\documentclass[12pt]{iopart}

\usepackage{setstack}
\usepackage{iopams}
\usepackage{amssymb}
\usepackage{array}
\usepackage{multirow}

\usepackage{natbib,har2nat}

\usepackage[pdfstartview=FitH,bookmarksopen=false,colorlinks,citecolor=blue]{hyperref}
\usepackage{graphicx}

\usepackage[dvipsnames]{xcolor}
\usepackage{xspace}

\newcommand{\figref}[1]{Figure \ref{#1}}
\newcommand{\secref}[1]{Section \ref{#1}}
\newcommand{\tabref}[1]{Table \ref{#1}}

\newcommand{\argmin}{\mathop{\mathrm{argmin}}}
\newcommand{\argmax}{\mathop{\mathrm{argmax}}}
\newcommand{\pixel}{\mathbf{x}}

\begin{document}

\title[]{Fully Automated Treatment Planning for Head and Neck Radiotherapy using a Voxel-Based Dose Prediction and Dose Mimicking Method} 

\author{Chris~McIntosh$^1$, Mattea Welch$^1$, Andrea McNiven$^{1,2}$, David A. Jaffray$^{1,2}$, Thomas~G.~Purdie$^{1,2}$}
\address{$^1$ Department of Medical Imaging \& Physics, Princess Margaret Cancer Centre, University Health Network (UHN), Toronto, ON, Canada}
\address{$^2$ Department of Radiation Oncology, University of Toronto, Toronto, ON, Canada}
\ead{\mailto{chris.mcintosh@rmp.uhn.on.ca},\mailto{tom.purdie@rmp.uhn.on.ca}}

\date{\today}

\begin{abstract}
Recent works in automated radiotherapy treatment planning have used machine learning based on historical treatment plans to infer the spatial dose distribution for a novel patient directly from the planning image. We present an atlas-based approach which learns a dose prediction model for each patient (atlas) in a training database, and then learns to match novel patients to the most relevant atlases. The method creates a spatial dose objective, which specifies the desired dose-per-voxel, and therefore replaces any requirement for specifying dose-volume objectives for conveying the goals of treatment planning. A probabilistic dose distribution is inferred from the most relevant atlases, and is scalarized using a conditional random field to determine the most likely spatial distribution of dose to yield a specific dose prior (histogram) for relevant regions of interest. Voxel-based dose mimicking then converts the predicted dose distribution to a deliverable treatment plan dose distribution. In this study, we investigated automated planning for right-sided oropharaynx head and neck patients treated with IMRT and VMAT. We compare four versions of our dose prediction pipeline using a database of 54 training and 12 independent testing patients. Our preliminary results are promising; automated planning achieved a higher number of dose evaluation criteria in 7 patients and an equal number in 4 patients compared with clinical. Overall, the relative number of criteria achieved was higher for automated planning versus clinical (17 vs 8) and automated planning demonstrated increased sparing for organs at risk (52 vs 44) and better target coverage/uniformity (41 vs 31). The novel dose prediction method with dose mimicking can generate deliverable treatment plans in 12-13 minutes without any user interaction. The method is a promising approach for fully automated treatment planning and can be readily applied to different treatment sites and modalities.
\end{abstract}

\maketitle 

\section{Introduction}
\label{sec::Intro}

Radiotherapy is appropriate for nearly $50\%$ of all cancer patients \cite{delaney2005role}. While radiotherapy is commonly used as a curative treatment, it can also be used as an adjuvant treatment to mitigate the chance of recurrence, a neo-adjuvant treatment intended to reduce the size of tumors prior to other cancer treatment modalities, or as a palliative treatment for disease control and symptom relief \cite{baskar2012cancer}. The variation in the use of radiotherapy creates obstacles to developing standardized treatment planning approaches and necessitates automated methods with specific considerations for each treatment site, prescription dose and intent.

Regardless of role, the development of radiotherapy treatment plans is a time consuming process involving oncologists, dosimetrists and medical physicists. In addition, there are many manual steps in the radiotherapy process and treatment planning often requires numerous iterations for plan optimization; a process that can take anywhere from a few hours to multiple days \cite{das2009analysis}, and results in inter- and intra-institutional variation in planning practice and quality \cite{nelms2012variation}. In many cases time constraints result in sub-optimal plans being delivered in the clinic \cite{nielsen2005audit}, resulting in worse patient outcomes \cite{abrams2012failure}, \cite{peters2010critical}. Additionally, with the emergence of adaptive radiation therapy \cite{juloori2015adaptive,ghilezan2010adaptive}, there is an increasing demand for high throughput planning while maintaining quality, creating a need for automation. 

The canonical radiotherapy planning process involves: 1) contouring regions of interest (ROI) on volumetric planning images, including target volumes, normal tissue organs-at-risk (OAR) volumes, in addition to non-anatomical ROIs strictly for the purposes of treatment planning; 2) defining dose-volume objectives based on dose prescriptions and dose constraints specified by targets and OARs intended to optimize the dose delivered by each beam; 3) setting up the beam geometry; 4) running an optimization engine to attempt to balance between the dose-volume objectives; and 5) calculating the final plan. A key challenge in the planning process is to determine when compromises need to be made from the ideally achieved clinical goals and provide the most relevant trade-off for competing clinical priorities.

Setting the dose-volume objectives to try and achieve the optimal trade-off for these competing priorities represents a significant manual step in the radiotherapy treatment planning process. Clearly this is a challenging problem because there is subjectivity in which trade-offs are made. Most automated planning methods try to determine dose-volume objectives automatically based on previously treated patients using a limiting number of patient/anatomical features drawn from ROIs \cite{kazhdan2009shape,appenzoller2012predicting,wu2013using,yang2013overlap,shiraishi2015knowledge}. There are two challenges to this approach. Firstly, features taken only from contoured structures can ignore other image appearance characteristics and may not be sufficient to model these competing priority details \cite{mcIntosh2016TMIAutoPlan}. Secondly, dose-volume objectives are cumulative distribution functions summed over the dose to a particular ROI, and are thus spatially insensitive within the ROI. Consequently, in cases where there is a desire to shape the dose within an ROI, or for tissue not defined by an ROI, additional planning ROIs and associated dose-volume objectives are created ad-hoc. These planning ROIs are used to spatially guide the dose, and are not evaluated during quality assurance. We seek to address these issues by instead using the atlas patients to directly infer the dose-per-voxel for the novel patient planning image based on patient geometry, ROI-based features, and both global and local image intensity features (texture, gradient, etc).

One treatment site that exemplifies the requirement for managing competing clinical priorities is head and neck. The proximity and intersection of OARs with multiple targets prescribed to up to three prescription dose levels is a challenging site for both manual and automatic radiotherapy treatment planning pipelines. There has been a desire to improve standardization and efficiency for head and neck due to the tremendous complexity and time required to generate clinically acceptable treatment plans \cite{peters2010critical,batumalai2013important,wu2013using}.

We have previously developed a novel contextual atlas regression forest (cARF) method that produces a probabilistic estimate of the dose at each voxel in the planning image as opposed to an estimate of the dose to a region, thus creating a spatial dose objective \cite{mcIntosh2016TMIAutoPlan,mcIntosh2016PMBAutoPlan}. The approach utilizes machine learning and radiomic image features to estimate the dose-per-voxel based on the dose to voxel-feature relationship observed in the most similar patients from a training database. Selection of the most similar patient is automatically learned based on the most relevant features to dose prediction. We extended the method to include a dose distribution prior (analogous to a DVH) for each target and OAR \cite{mcIntosh2016PMBAutoPlan}. This approach combines the dose-per-voxel probability distribution with dose-volume objective based approaches, finding the most likely spatial distribution of dose (based on learning from training atlases) under an inferred prior on the distribution frequency (analogous to a DVH). The dose prior used is a joint distribution prior over all targets and OARs in a plan, thus encoding the particular trade-off of any atlas-patient as a function of the intersections that necessitate the most challenging trade-offs. The previous study explored both our ARF algorithm \cite{mcIntosh2016PMBAutoPlan} and deformable atlas registration for mapping the dose distribution from an atlas patient onto the novel patient, as well as multiple atlas-selection approaches including an atlas-free method, overlap volume histograms \cite{Kazhdan2009}, neighborhood approximation forests \cite{konukoglu2012neighbourhood}, and our proposed cARF. Our method achieved the highest overall accuracy, and is the focus of this work. Similarly, Shiraishi and Moore developed an artificial neural network approach for voxel-based dose prediction \cite{shiraishi2016knowledge}, which uses one trained model for all patients instead of an atlas-selection approach. As we noted  \cite{mcIntosh2016TMIAutoPlan,mcIntosh2016PMBAutoPlan}, while these methods provide predictions of the dose distribution, and hence the spatial dose and dose-volume objectives, the predictions themselves are made per-voxel (or ROI) and are thus not necessarily clinically achievable dose distributions that consider beam geometry, scatter, and attenuation. In this work we use dose mimicking to examine the clinical deliverability of these plans.

This work is the first to convert a predicted per-voxel dose distribution into a deliverable radiotherapy treatment plan.  Multiple patient atlas selection and machine learning methods are used to predict per-voxel dose distributions. We integrate these methods with dose mimicking to create a novel fully automated radiotherapy treatment planning pipeline. The automated planning pipeline only requires the target and a limited number of anatomical OARs as input, and the pipeline does not require any dose-volume objectives to be specified. This preliminary study is also the first work to apply voxel-based dose prediction to the head and neck treatment site and to determine the applicability of voxel-based dose prediction for automated treatment planning scored against dose evaluation criteria.

\section{Methods and Materials}
\label{sec::methods}
\subsection{Patients}
In this work, the contextual atlas regression forest (cARF) framework was trained and validated using 66 right-sided oropharynx head and neck patients planned and treated between 2011 and 2013 according to our institutional protocol. For analysis the data was divided into: 54 training image-plan pairs; and a disjoint set of 12 testing instances. The 12 patients for treatment planning were randomly selected from the database of 66 patients with no exclusion criteria.  

For all patients, clinical treatment plans were generated with the Pinnacle$^3$ treatment planning system (Philips, Madison, WI) and treated on the Varian TrueBeam (6 MV, 120 leaf multi-leaf collimator). The data consisted of 59 intensity modulated radiation therapy (IMRT) plans and 7 volumetric modulated arc therapy (VMAT) plans. The IMRT plans were delivered using a 9 beam step-and-shoot beam arrangement with approximately equi-distant beam spacing starting anteriorly. The VMAT plans were delivered using two 330 degree arcs with 3 degree gantry spacing between control points. 

All patients were treated in 35 fractions and prescribed 7000 cGy to the high risk target volume (gross disease and involved nodes) and 5600 cGy to the elective target volume (bilateral nodal neck region) (\tabref{tb::patientsummary}). 7 patients were also prescribed 6300 cGy to an intermediate-risk target volume (suspicious nodes) and 5 patients had the 7000 cGy high risk target volume on both the right and and left side. Following our clinical protocol, all delineated clinical target volumes (CTV) are expanded by a uniform 5 mm margin to create corresponding planning target volumes (PTV). For optimization and evaluation purposes, PTVs were then contracted from the external contour of the patient by 5 mm. Target volumes were delineated by the treating radiation oncologist and all target delineations were reviewed in peer-review case rounds. OARs were delineated by the radiation oncologist and/or treatment planners.

\newcolumntype{x}[1]{>{\centering\arraybackslash\hspace{0pt}}p{#1}}
\begin{table}
\caption{Summary of Patients Evaluated}
	\label{tb::patientsummary}
\footnotesize
\lineup
\begin{tabular*}{\linewidth}{@{}lllx{55pt}x{55pt}x{100pt}}
\br
\textbf{}&& \textbf{High Risk} & \centre{3}{\underline{\hspace{30pt}\textbf{Planning Target Volume (cm$^3$)}\hspace{30pt}}}\\
\textbf{Patient \#} & \textbf{Technique} & \textbf{Target Location} & \textbf{High Risk} & \textbf{Elective} & \textbf{Intermediate Risk} \\
\mr
1 & VMAT & Right, Left & \textbf{388.5} & \textbf{797.7} & \textbf{---}\\
2 & IMRT & Right &  \textbf{187.8} & \textbf{712.2} & \textbf{61.2}\\ 
3 & IMRT & Right & \textbf{360.4} & \textbf{1187.8} & \textbf{23.3}\\
4 & IMRT & Right & \textbf{360.6} & \textbf{914.6} & \textbf{34.4}\\
5 & IMRT & Right & \textbf{238.3} & \textbf{918,8} & \textbf{40.7}\\
6 & VMAT & Right & \textbf{187.1} & \textbf{785.3} & \textbf{---}\\
7 & IMRT & Right, Left & \textbf{88.3} & \textbf{513.5} & \textbf{---}\\
8 & IMRT & Right, Left & \textbf{313.0} & \textbf{673.0} & \textbf{---}\\
9 & IMRT & Right & \textbf{244.8} & \textbf{871.2} & \textbf{38.2}\\
10 & IMRT & Right & \textbf{173.5} & \textbf{867.0} & \textbf{136.6}\\
11 & IMRT & Right, Left & \textbf{296.2} & \textbf{885.3} & \textbf{55.5}\\
12 & IMRT & Right, Left & \textbf{637.0} & \textbf{698.8} &\textbf{---} \\

\br
\end{tabular*}
\newline
\begin{tabular}{@{}ll}

IMRT: Intensity Modulated Radiation Therapy \\
VMAT: Volumetric Modulated Arc Therapy \\
High Risk: Gross disease and involved nodal volume precribed to 7000 cGy\\
Elective: Bilateral nodal neck volume precribed to 5600 cGy\\
Intermediate Risk:Suspicious nodal volume prescribed to 6300 cGy\\

\end{tabular}
\end{table}

\subsection{Automated spatial dose prediction pipeline}
\begin{figure*}
	\begin{center}
		\begin{tabular}{c}
        \includegraphics[width=\linewidth,clip=]{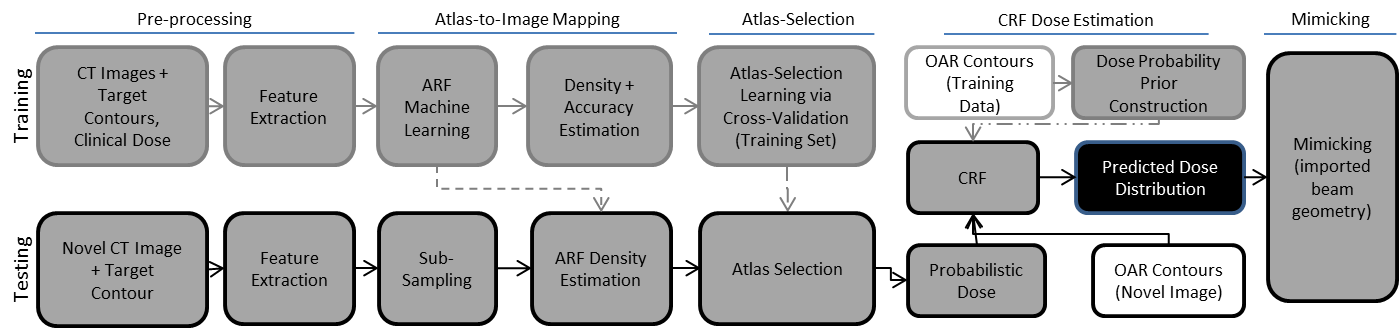}\\
		\end{tabular}
	\end{center}
	\caption[]{Flow chart showing training and testing pipelines of proposed voxel-based dose prediction algorithm with dose mimicking using collapsed cone convolution dose calculation. Dashed lines indicate learned output models from the training phase being used to predict dose for novel images. Figure adapted from \cite{mcIntosh2016PMBAutoPlan} with permission.}
    \label{fig::flowChart} \end{figure*}

The proposed planning pipeline predicts a probabilistic dose estimate at each image voxel using the cARF framework. The prediction pipeline uses gradient-based image features to characterize patient geometry, ROI shape, and both global and local image appearance. Regression is used to classify the feature and radiation dose relationship, and the most relevant atlases are automatically selected from an image database using density estimation over observed features. The pipeline consists of two learning phases. In the atlas-to-image mapping phase, atlas regression forests (ARF) are trained to model the relationship between a patient’s image features and their clinical dose distribution per-voxel. In the atlas-selection phase, a second model is trained to select the most relevant ARFs for a novel patient image based on density estimation over observed image features. At a high level the process is analogous to an expert looking at a novel image, searching a database for the most similar patient (the atlas), and then mapping the dose distribution from the atlas to the novel patient. Finally, a conditional random field model is optimized to find the most probable spatial distribution that conforms to a joint prior distribution over dose values for the ROIs. An overview of the process is presented in \figref{fig::flowChart}, and complete details of the atlas-to-image mapping, and atlas-selection algorithm are presented in \cite{mcIntosh2016TMIAutoPlan}, with an overview and details of the conditional random field dose prior in \cite{mcIntosh2016PMBAutoPlan}. In what follows we provide a relevant summary.

The first learning phase involves a training data set composed of patient images and clinical dose distribution pairs. A texture filter bank is used to compute a set of image features for each voxel, and defined for each pair during this phase. Optionally this also contains patient specific planning ROIs, as well as target and organ position and geometry. The relationship between the image feature set and the clinical dose distribution is then modeled as a regression problem using an ARF on a voxel-by-voxel basis, taking into consideration the contextual image information. Each training pair includes a training image $I(\pixel)$ and its corresponding clinical dose $d(\pixel)$. ARFs are extensions of random forests \cite{Breiman2001}, which are large collections of generalized decision trees (regression models) where the votes from each tree is counted to create a probabilistic output. Each ARF takes as input a vector of image features, $F_{\pixel}$, for voxel $\pixel$, and outputs a probabilistic estimate of the dose, $d$: $P(d_{\pixel}|F_{\pixel})$ at $\pixel$. 

Following \cite{mcIntosh2016TMIAutoPlan}, feature extraction uses both non-contoured and contoured image data. Features are calculated across a range of scales via convolution with a 3D texture filter bank that is an extension of the Leung-Malik filter bank \cite{leung2001}. We use a total of $86$ filters including $84$ first and second order derivatives parameterized across $14$ rotations and three scales $\sigma=\{24,48,64\}$, and two rationally invariant filters in the form of isotropic Gaussians and Laplacian of Gaussians. Filter parameters remain fixed from \cite{mcIntosh2016PMBAutoPlan}, selected through manual iteration on the training data which is completely independent from both the training and testing data used for this study, as it did not include head and neck cases. To account for variable voxel spacing during RT planning image acquisition the filtering is performed in millimetres in world co-ordinates. We also include 4 target specific features for each target: a signed distance transform of the target $\Phi_{\pixel}$, and a vector in 3D denoting the direction and distance to the closest point on the target boundary:
$\left(x-c\mid c = \argmin_{c} \parallel x-c {\parallel}_2 \forall c \in C_1\right)$, where $C_1$ is the first target, for example. In our previous studies we did not include features from the OARs, in our experiments in this study we compare our proposed pipeline with and without those additional features.

In our cARF model we further use a density estimate that models the likelihood of observing a feature given the image $P(F|I)$, learned during the ARF training process from the features selected by the ARF for dose prediction. The density estimates are statistical representations of global image appearance, modeling the frequency at which the features (e.g. lung like image patch) occur. Finally in this phase we also learn a joint dose distribution prior $P(d|R(\pixel))$ where $R(\pixel):\Omega \mapsto \{0,1\}^{|\mathcal{C}|}$ is a binary encoding of voxel-wise ROI membership from the image domain, $\Omega$, onto the set of ROIs, $\mathcal{C}$. Note that $P(d|R(\pixel))$ is a joint prior over all ROIs and their unique intersections in contrast to a typical DVH, which is a cumulative prior distribution per ROI.

The second phase of the pipeline involves learning to select the most similar ARFs for a novel image by measuring the difference between the learned density estimate and the observed features from the novel image. Density estimates are chosen since they enable image similarity estimation as a function of the features observed to be important to dose prediction. The density estimation learns, for example, that a particular training image having a target near an OAR was a key feature in generating its dose distribution. Thus, for a novel patient with a target near the OAR, only ARFs from patients with similar targets near the OAR are selected and used for dose prediction. Leave-one-out cross-validation is used to train a model to predict the accuracy of an ARF for a novel patient given the density differences. First, each ARF is used to predict the dose for all other training images, and the resulting dose prediction accuracy is measured. Then a random forest model is trained to predict the accuracy for each image using the density differences as the features, and the dose prediction accuracy as the regression variable. The key is that these models can then predict the accuracy of an ARF for a novel image based on image features alone, i.e. without knowing the dose distribution of the novel image. 

After both learning phases are complete, the pipeline is validated using novel patient images. The feature set is calculated for each novel image and the most similar ARFs are selected as those with the least distance. These ARFs are then used to predict a probabilistic model of the dose distribution. An average dose prior is also computed over the $k$ most similar models, $P(d|R(\pixel))$. Finally, as presented in \cite{mcIntosh2016PMBAutoPlan}, a conditional random field (CRF) model \cite{Lafferty2001} is used to find the most likely spatial assignment of dose-per-voxel according to $P\left(d_{\pixel}|F_{\pixel}\right)$ while adhering to the dose prior. The resulting optimization problem can be written as a CRF:

\begin{equation}
\label{eq::mapEstimatePrior}
\tilde{d} =  \argmax_{d_{\pixel} \in \Omega} \prod_{\pixel} P\left(d_{\pixel}| F_{\pixel}\right)P\left(d|R(\pixel)\right),
\end{equation}

\noindent and optimized as a binary linear program. The CRF is optimized to obtain a scalar dose estimate that conforms to a predicted joint dose distribution prior from the most similar atlases. 

\subsection{Effects of additional OAR ROIs}
In this work we experiment with including additional OAR ROI features. The ROIs included in this version of the pipeline consisted of the high risk PTV, elective PTV, intermediate risk PTV (if delineated), as well as the esophagus, left parotid, and right parotid. The CRF modeled dose priors for these ROIs, and additionally the mandible, larynx, postcricoid, spinal cord, and brainstem.  

\subsection{Dose Mimicking}
\label{sec::doseMimicking}
We used voxel-based dose mimicking to convert the predicted cARF dose distributions and the original clinical dose distribution into new deliverable radiotherapy treatment plans \cite{Petersson2016}. The dose mimicking was compromised of three intermediate collapsed cone convolution dose calculations over the course of multiple dose mimicking iterations and then a final collapsed cone convolution dose calculation representing the final plan dose distribution for each method.

The predicted cARF dose distributions and the original clinical dose distributions were imported into RayStation (version 4.5, RaySearch Laboratories, Stockholm, Sweden) for dose mimicking. The beam arrangement for the clinical plans was used for the dose mimicking step for all the methods. The IMRT plans were mimicked using 9 step and shoot beams and the VMAT plans were mimicked using two partial arcs (330 degrees) with a 3 degree control point spacing (101 control points per arc). 

The dose mimicking objectives used in the current study are voxel-based, therefore dose mimicking matches dose at each voxel over the entire dose grid, without considering the dose as a distribution per ROI i.e. dose volume histogram, or any dose evaluation objectives. The same set of spatial dose mimicking objectives and weights were used for all patients and methods. The objectives and weights were established by mimicking all of the original clinical dose distributions to determine a single set of objectives and weights that provided good agreement between dose distributions pre- and post- dose mimicking across all patients. The same set of dose mimicking objectives and weights set from the clinical dose mimicking phase were then applied to the dose mimicking of predicted cARF dose distributions to generate the final cARF treatment plans.

\section{Results}
\label{sec::results}

\begin{table}[ht!]
\caption{Summary of Dose Evaluation Criteria}
	\label{tb::criteriaSummary}
\footnotesize
\lineup
\begin{indented}
\item[]\begin{tabular}{llll}
\br
Label & ROI & Evaluation & Limit (cGy)\\
\mr
\bs
BrStem Dmax & Brainstem & Dose at 0.1 cm$^3$ & $<5000$ \\ 
Chiasm Dmax & Optic Chiasm & Dose at 0.0 cm$^3$ & $<5200$ \\
Cord Dmax& Spinal Cord & Dose at 0.1 cm$^3$ & $<4500$ \\ 
Esoph DMax & Esophagus & Dose at 0.1 cm$^3$ & $<5000$ \\
Larynx D67\% & Larynx & Dose at 67\% volume & $<5000$ \\ 
LParotid D50\% & Left Parotid & Dose at 50\% volume & $<3000$ \\
LPlexus Dmax & Left Brachial Plexus & Dose at 0.0 cm$^3$ & $<6300$ \\ 
Mandible Dmax & Mandible & Dose at 0.1 cm$^3$ & $<7350$ \\ 
PTV56 Dmax & Elective PTV to 5600 cGy & Dose at 0.0 cm$^3$ & $<7000$ \\
PTV56 D99\% & Elective PTV to 5600 cGy & Dose at 99\% volume & $>5320$ \\ 
PTV56 D95\% & Elective PTV to 5600 cGy & Dose at 95\% volume & $>5600$ \\ 
PTV70 Dmax & High Risk PTV to 7000 cGy & Dose at 0.0 cm$^3$ & $<8050$ \\
PTV70 D99\% & High Risk PTV to 7000 cGy & Dose at 99\% volume & $>6650$ \\ 
PTV70 D95\% & High Risk PTV to 7000 cGy & Dose at 95\% volume & $>7000$ \\ 

\br
\end{tabular}
\end{indented}
\end{table}

We evaluated four versions of our method: the cARF method without and with OAR features denoted as cARF and cARF[ROI] respectively; and each of those methods without and with the CRF dose prior optimization, denoted as cARF-CRF and cARF-CRF[ROI] respectively. As previously noted, all cARF algorithm parameters were set to the previously obtained values used in \cite{mcIntosh2016PMBAutoPlan}. For brevity, the  collapsed cone convolution calculated dose distributions for the final treatment plan will simply be referred to by the method name and the original clinical or predicted cARF dose distributions will be explicitly indicated.

\begin{table}[t]
\caption{Summary of Total Achieved Dose Evaluation Criteria}
	\label{tb::resultSummary}
\footnotesize
\lineup
\begin{tabular}{llllllllllll}
\br
 & \centre{2}{cARF} & \centre{2}{cARF-CRF} & \centre{2}{cARF[ROI]} & \centre{2}{cARF-CRF[ROI]} & \centre{2}{Clinical} \\
Patient \# & Pred. & Plan & Pred. & Plan & Pred. & Plan & Pred. & Plan & Original & Plan \\
\ns\ns\ns
&\crule{2}&\crule{2}&\crule{2}&\crule{2}&\crule{2}\\
1 & 12 & 11 & 11 & 11 & 12 & 9 & 11 & 12 & 11 & 11 \\
2 & 10 & 11 & 12 & 11 & 10 & 10 & 12 & 12 & 10 & 11 \\
3 & 12 & 11 & 11 & 11 & 12 & 11 & 11 & 11 & 11 & 11 \\
4 & 11 & 8 & 11 & 11 & 12 & 9 & 10 & 11 & 11 & 11 \\
5 & 11 & 12 & 12 & 13 & 13 & 13 & 12 & 13 & 13 & 13 \\
6 & 12 & 13 & 13 & 14 & 13 & 14 & 12 & 12 & 12 & 14 \\
7 & 10 & 10 & 12 & 12 & 10 & 11 & 11 & 12 & 10 & 11 \\
8 & 9 & 11 & 11 & 11 & 10 & 12 & 11 & 12 & 12 & 12 \\
9 & 11 & 12 & 12 & 11 & 12 & 11 & 11 & 12 & 10 & 10 \\
10 & 13 & 12 & 12 & 12 & 13 & 13 & 12 & 13 & 10 & 10 \\
11 & 11 & 10 & 11 & 10 & 10 & 9 & 11 & 10 & 9 & 9 \\
12 & 10 & 7 & 11 & 9 & 9 & 10 & 11 & 10 & 8 & 8 \\
\ns
&\crule{10}\\
Total: & 132 & 128 & 139 & 136 & 136 & 132 & 135 & 140 & 127 & 131\\
\br
\end{tabular}
\newline
\begin{tabular}{@{}ll}
Pred.: & Predicted spatial dose distribution without dose mimicking\\
Original: & Original clinical plan dose distribution without dose mimicking\\ 
Plan: & Final treatment plan dose distribution after dose mimicking \\
cARF: & Contextual Atlas Regression Forests  \\
CRF: & Conditional Random Field \\
\end{tabular}
\end{table}
    
\begin{figure}
	\begin{center}
		\begin{tabular}{c}
         \includegraphics[width=0.8\linewidth,clip=]{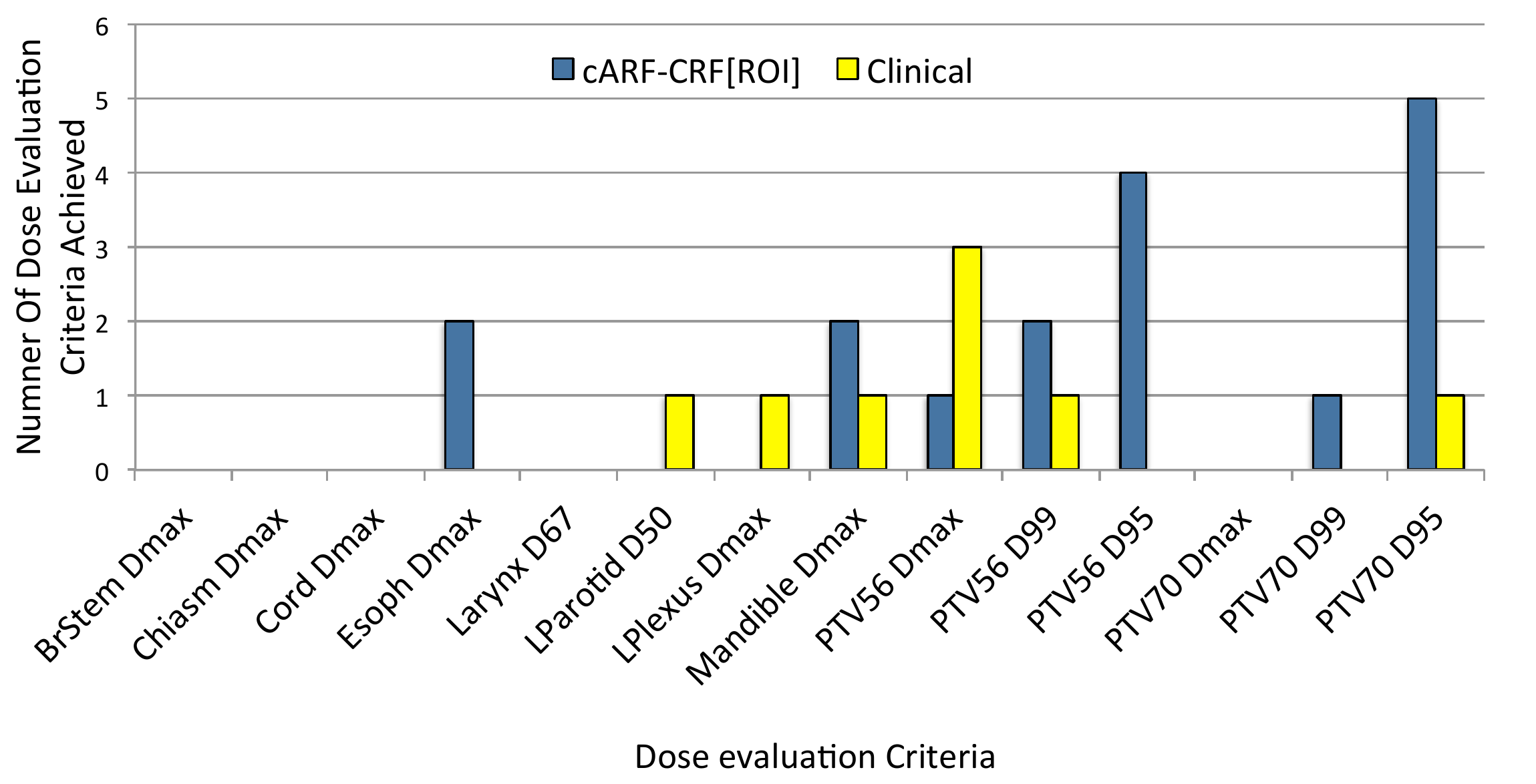}\\
         (a)\\
        \includegraphics[width=0.8\linewidth,clip=]{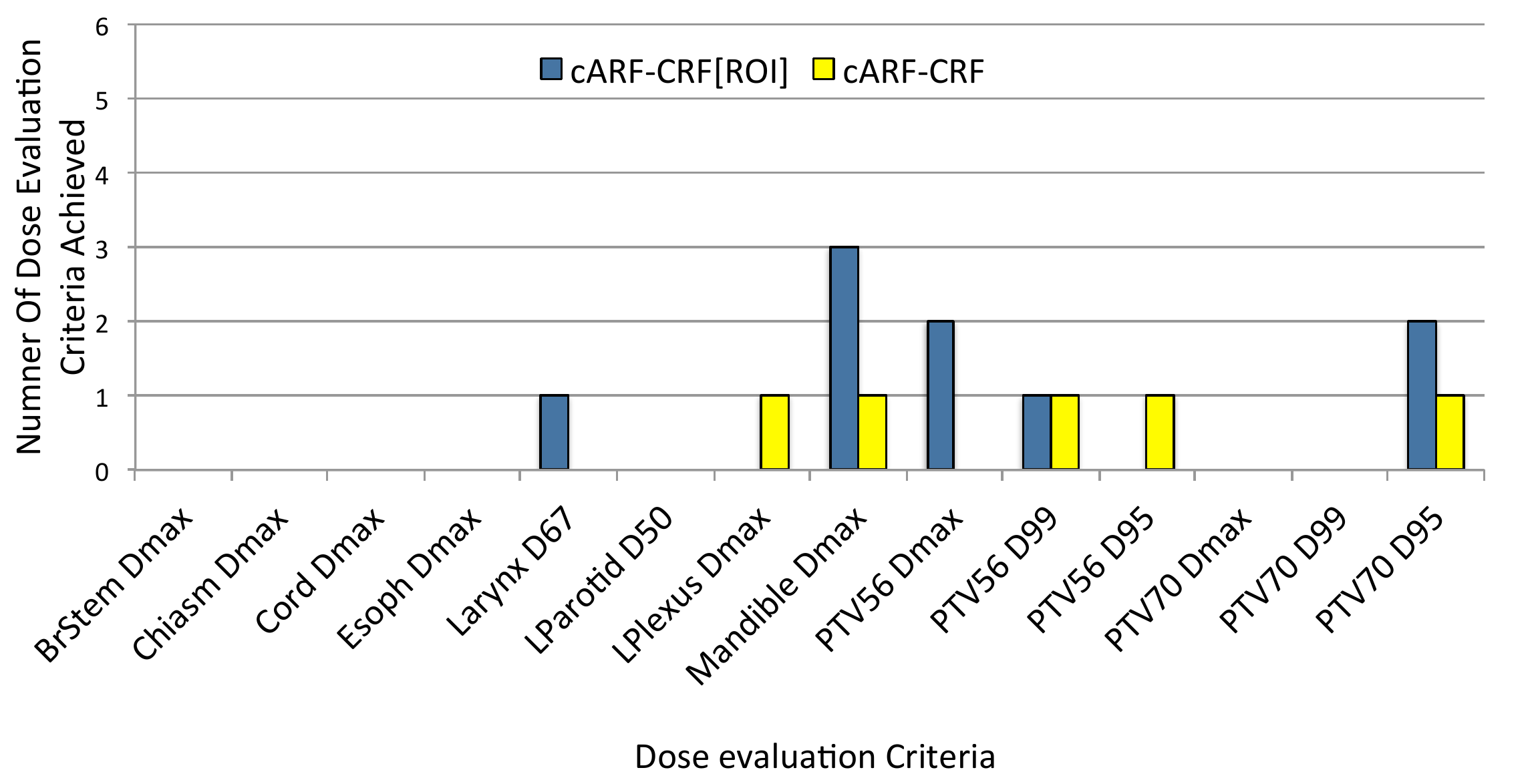}\\
        (b)\\
		\end{tabular}
	\end{center}
	\caption{The relative number of dose evaluation criteria uniquely achieved per patient for cARF-CRF[ROI] dose distributions compared to a) clinical dose distributions and b) cARF-CRF dose distributions as a function of the dose evaluation criteria used in the study. For example, the cARF-CRF[ROI] dose distributions achieved the PTV56 D99 criteria for 2 patients which the clinical failed, but failed in one patient that the clinical achieved. Overall, the cARF-CRF[ROI] dose distribution achieved more dose evaluation criteria compared with clinical (17 versus 8) and compared with cARF-CRF (9 versus 5).}
    \label{fig::CriteriaDiff}
\end{figure}
    
\begin{figure}
	\begin{center}
		\begin{tabular}{c}
        \includegraphics[width=0.8\linewidth,clip=]{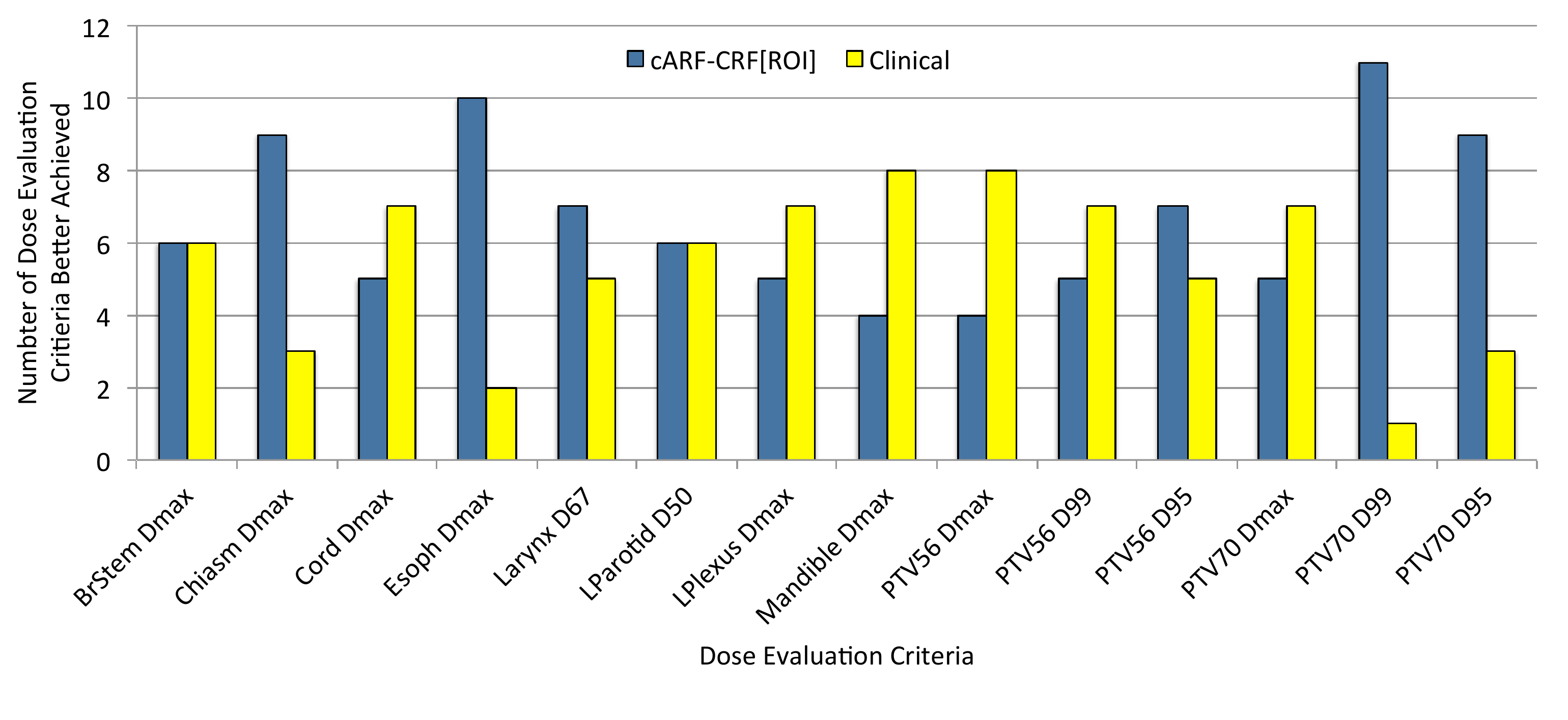}\\
        (a)\\
        
        \includegraphics[width=0.8\linewidth,clip=]{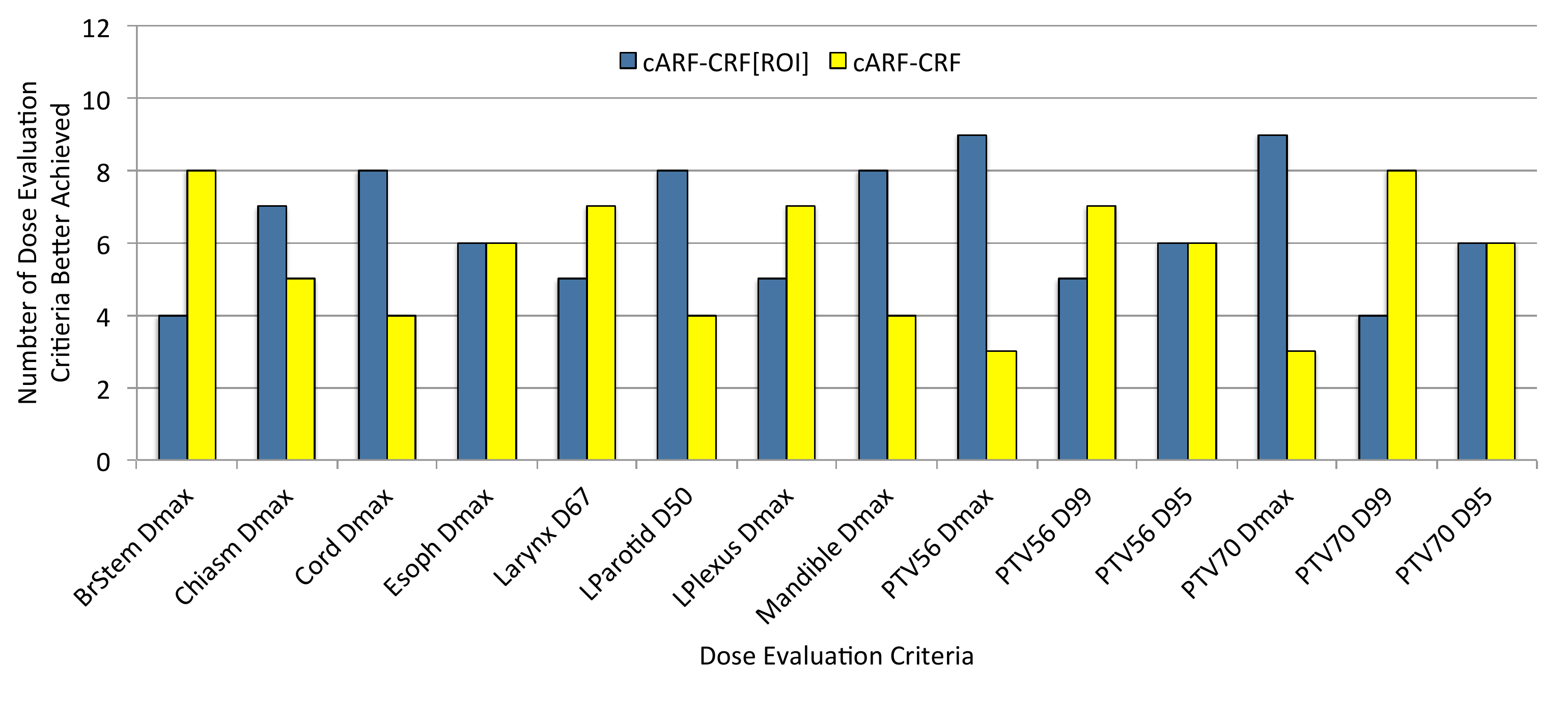}\\
        (b)\\
		\end{tabular}
	\end{center}
	\caption{The number of dose evaluation criteria that were better achieved per patient for cARF-CRF[ROI] dose distributions compared to a) clinical dose distributions and b) cARF-CRF dose distributions as a function of the dose evaluation criteria used in the study. The method with the lower dose for sparing criteria or higher dose for target coverage criteria per patient was counted. For example, the minimum dose for the PTV70 D99 dose evaluation criteria was higher in 11 patients for the cARF-CRF[ROI] dose distribution and lower in one patient compared with the clinical dose distribution. Overall, the cARF-CRF[ROI] dose distribution better achieved dose evaluation criteria versus clinical (93 versus 75) and cARF-CRF (90 versus 78 cases).}
\label{fig::DoseDiff}
\end{figure}

Dose distributions were scored against our institutional dose evaluation criteria (\tabref{tb::criteriaSummary}), and summary results are presented for each method and for each patient (\tabref{tb::resultSummary}). The criteria were scored as a binary pass or fail, and the score for each patient was summed over all 14 criteria. The dose evaluation criteria compliance for each imported dose distribution (predicted cARF and original clinical) and final plan dose distribution show that the cARF-CRF[ROI] method performed the best overall in terms of achieving dose evaluation criteria (140) although the cARF-CRF (136) and cARF[ROI] (132) methods also achieved more criteria than clinical (131) overall. The dose mimicking was able to improve the number of achieved criteria for the cARF-CRF[ROI] (135 to 140) and clinical (127 to 131) methods. However, the other cARF methods generated predicted dose distributions which were not as well mimicked, where some dose evaluation criteria that were achieved in the prediction were ultimately not achieved in the final plan. At the patient level, the cARF-CRF[ROI] achieved more dose evaluation criteria than clinical in 7 patients, achieved less criteria in 1 patient and met the same number of dose evaluation criteria in 4 patients. Similarly, the cARF-CRF achieved more dose evaluation criteria in 5 patients, less in 1 patient and tied in 6 patients compared with clinical.

The number of dose evaluation criteria uniquely achieved per patient by one method in comparison to another were scored for the cARF-CRF[ROI] dose distributions compared with the clinical and cARF-CRF dose distributions (\figref{fig::CriteriaDiff}). Each method received a point for each criteria passed that the competing method failed, and ties were ignored; i.e, dose evaluation criteria that were both either achieved or failed for a given patient in each comparison were not scored. This result highlights where and how the methods disagreed. Overall, the cARF-CRF[ROI] dose distribution uniquely achieved more dose evaluation criteria compared with clinical (17 versus 8) and compared with cARF-CRF (9 versus 5).

Dose distributions were also scored for target coverage and OAR sparing by comparing the dose difference at each dose evaluation criteria level (\figref{fig::DoseDiff}). The method that better achieved the given dose evaluation criteria for each patient received a point (there will always be a total of 12 points per dose evaluation criteria corresponding to 12 patients evaluated) for each of the comparisons. The results show that cARF-CRF[ROI] was able to achieve better dose at each of dose evaluation criteria level compared with clinical (93 versus 75). Specifically, the cARF-CRF[ROI] fared better compared to clinical on both targets (52 versus 44) and OARs (41 versus 31). From the results it can be seen that the cARF-CRF[ROI] plans more consistently achieve better coverage of the PTV70 target dose evaluation criteria and also better spare the optic chiasm and esophagus. The cARF-CRF[ROI] also performed better than the cARF-CRF overall (90 versus 78), and specifically for targets (39 versus 33) and OARs (51 versus 45).

The run-time of the complete automated planning pipeline varies as a function of the number of CT slices, the size of the dose grid, and the number of voxels comprising the ROIs. Testing of the dose inference pipeline was performed in Matlab on a compute server with two 8-core Intel Xeon processors clocked at 2.6 Ghz and 256 GB of system memory. The average cARF-ROI run-time for the 12 patients evaluated was approximately 10 minutes each, including loading the data, calculating the features, performing atlas-selection, dose prediction, and running the CRF model. Calculating the additional ROI features in the cARF[ROI] method variants is negligible. The average run-time for dose mimicking including multiple collapsed cone convolution dose calculations was approximately 2 minutes for IMRT plans and approximately 3 minutes for VMAT plans on the standard clinical treatment planning system hardware.

\begin{figure*}
	\begin{center}
        \includegraphics[width=0.9\linewidth,clip=]{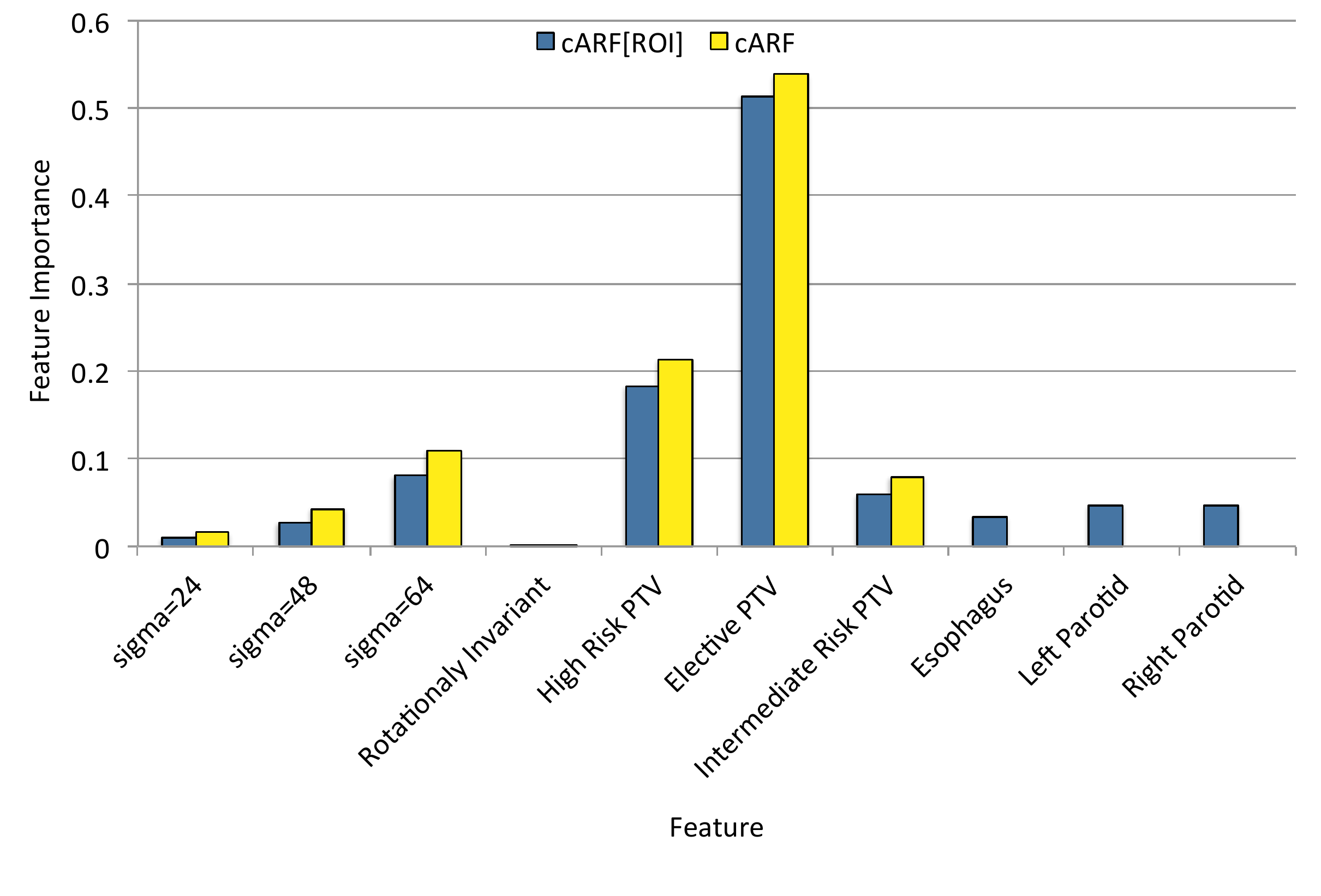}
	\end{center}
	\caption{Relative feature importance for voxel-based dose prediction for groups of related features with and without the additional OAR features. Features are grouped by type, including filtering features at multiple scales, $\sigma$, the rotationally invariant features, target, and OAR features.}
\label{fig::featureImp}
\end{figure*}
    
    We also evaluated the feature importance for the ARFs with and without OAR features (\figref{fig::featureImp}). The relative feature importance of each feature group, e.g. all features at a specific scale, is calculated using the out-of-bag samples during ARF training using the method detailed in \cite{Breiman2001}. Each forest in the ARF is trained on a different subset of voxels from a training image, and the remainder of the samples are the out-of-bag samples. Each feature in those samples is randomly permuted, and the impact the permutation has on the prediction accuracy is measured. Features with a larger impact are measured to have higher importance, since distorting those features greatly reduces accuracy. Without features from the OARs the algorithm turns to larger-scale gradient features and additional target features.

\section{Discussion and Related Work}
\label{sec::discussion}
\par 
As previously noted a number of works have explored knowledge-based automated planning \cite{kazhdan2009shape,appenzoller2012predicting,wu2013using,yang2013overlap,shiraishi2015knowledge,shiraishi2016knowledge}. The focus has been on using shape descriptors, e.g. the overlap volume histogram \cite{kazhdan2009shape}, to measure similarity between patients and directly infer resulting dose-volume objectives from the atlas database \cite{appenzoller2012predicting,wu2013using,yang2013overlap}. After specifying the dose-volume objectives, planning proceeds as normal. More recently, independent work by Shiraishi and Moore \cite{shiraishi2016knowledge}, and simultaneously our group \cite{mcIntosh2016TMIAutoPlan} has introduced the notion of predicting the dose on a voxel-by-voxel basis. We discussed our work in the introduction.  Using a database of atlas patients Shiraishi and Moore trained two dose prediction models: one inside the target and one outside. They validated their method on the spatial prediction of dose distributions for 12 prostate and 23 stereotactic radiosurgery plans, demonstrating improved OAR sparing in comparison to DVH models \cite{appenzoller2012predicting,shiraishi2015knowledge}. However, as we noted in \cite{mcIntosh2016TMIAutoPlan,mcIntosh2016PMBAutoPlan} these dose distributions are not necessarily clinically deliverable, even though they are derived from previously deliverable treatment plans. The current work, therefore completes the entire automated planning pipeline and generates clinically deliverable treatment plans based on a predicted dose distribution. In particular, we investigate the applicability of voxel-based dose prediction for automated head and neck radiotherapy treatment planning.

The 12 randomly selected patients exhibited a very good range in terms of target location, target size and target type (\tabref{tb::patientsummary}) and also high variability in plan complexity based on the number of dose evaluation criteria achieved in the clinical plans (\tabref{tb::resultSummary}). The majority of the clinical plans failed 3 dose evaluation criteria i.e. 11 criteria passed, with a range from 0 to 6 criteria failures. The majority of these failures were the PTV70 D95, PTV56 D95, and esophagus Dmax failing in 8, 6, and 9 patients, respectively; result not included in the tables. The cARF-CRF[ROI] method performed consistently well across all of the patients when considering patient variability and plan complexity. The most challenging cases as indicated by the number of clinical dose evaluation criteria failures were: a relatively large high risk target volume on both the right and left side with a large elective target volume and intermediate risk volume (Patient 11), and a very large right-sided high risk target volume (Patient 12). However, it is interesting that other patients (Patients 2-5) had a similar presentation to Patient 11, and the cARF-CRF[ROI] plan achieved similar criteria that the clinical plan struggled with.

\begin{figure*}
\footnotesize
	\begin{center}
		\begin{tabular}{@{}cc}
         	\includegraphics[width=0.4\linewidth,clip=]{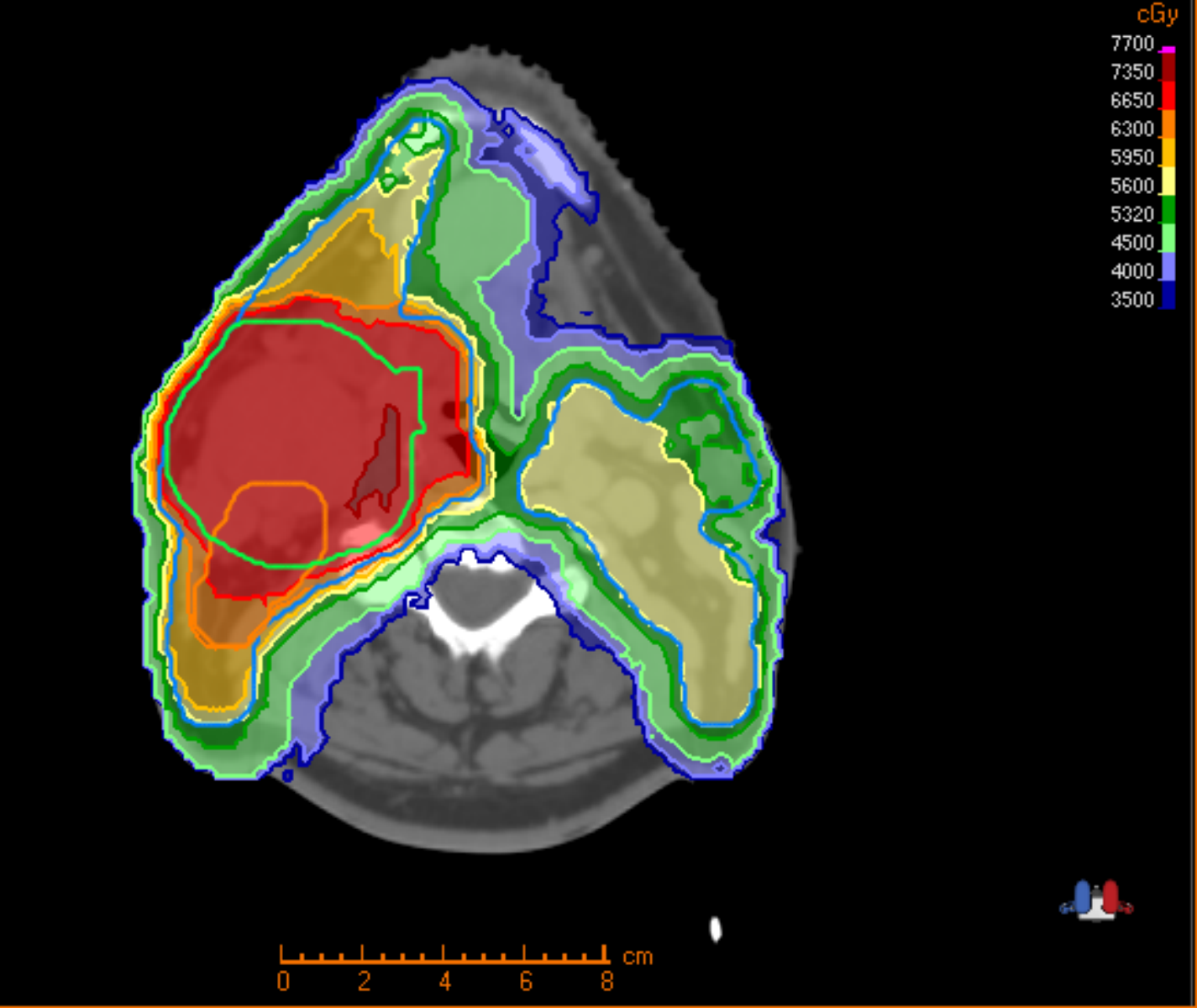} &
         	\includegraphics[width=0.4\linewidth,clip=]{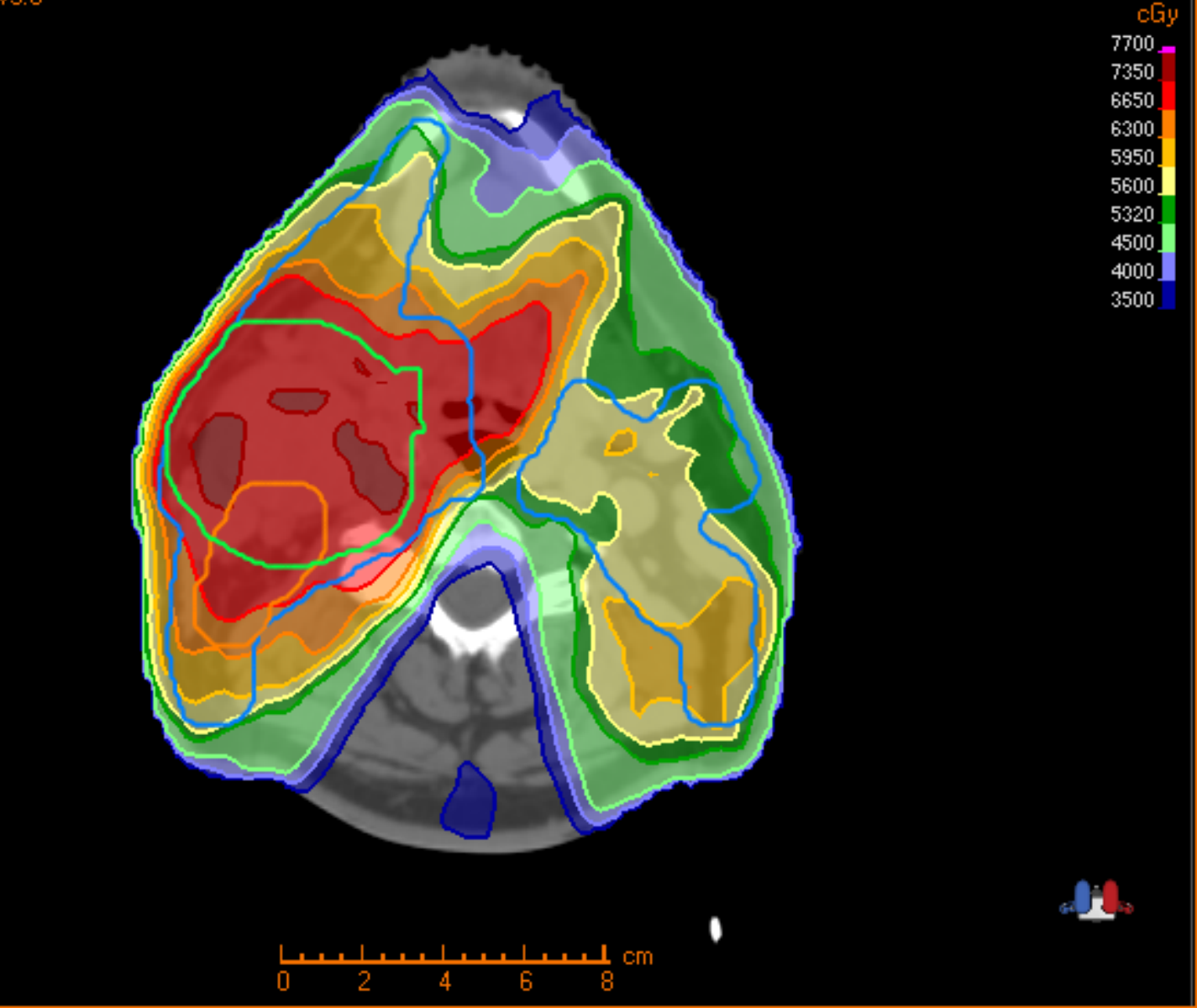}\\
         Predicted cARF-CRF[ROI] & cARF-CRF[ROI] Plan \\
		\end{tabular}
        \begin{tabular}{@{}cc}
        	\includegraphics[width=0.4\linewidth,clip=]{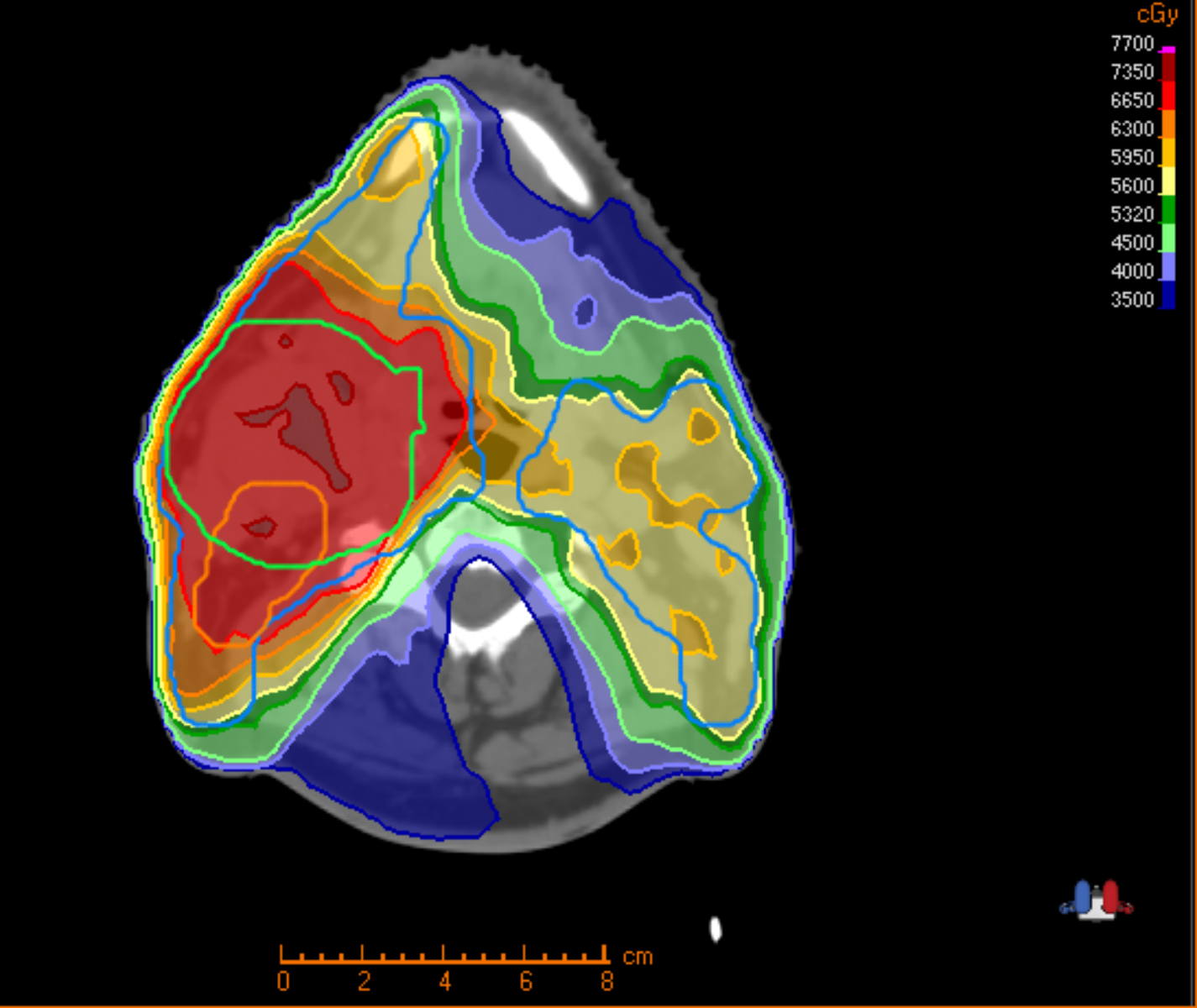} & 	  	             			 \includegraphics[width=0.4\linewidth,clip=]{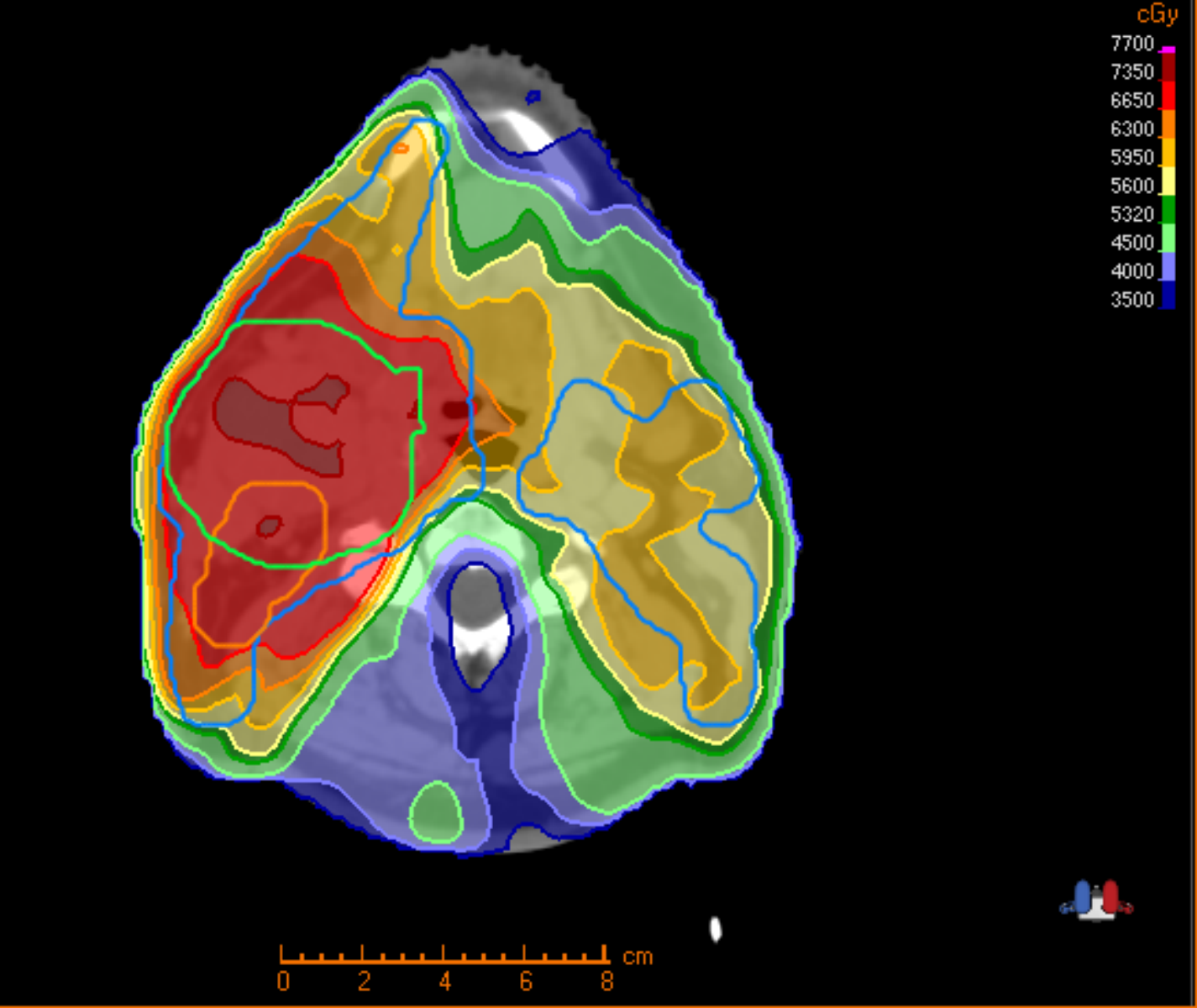}\\
             Original Clinical & Clinical Plan \\
		\end{tabular}

	\end{center}
	\caption[]{Pre- and post- dose mimicking axial dose distributions for cARF-CRF[ROI] and clinical. The dose distributions are for Patient 5 showing the high risk PTV prescribed to 7000 cGy (green line), elective PTV prescribed to 5600 cGy (light blue line), and intermediate risk PTV prescribed to 6300 cGy (orange line).}
    \label{fig::PrePostMimic} 
\end{figure*}

\begin{figure*}
	\begin{center}
    \begin{tabular}{@{}cl}
         \includegraphics[width=0.25\linewidth,clip=]{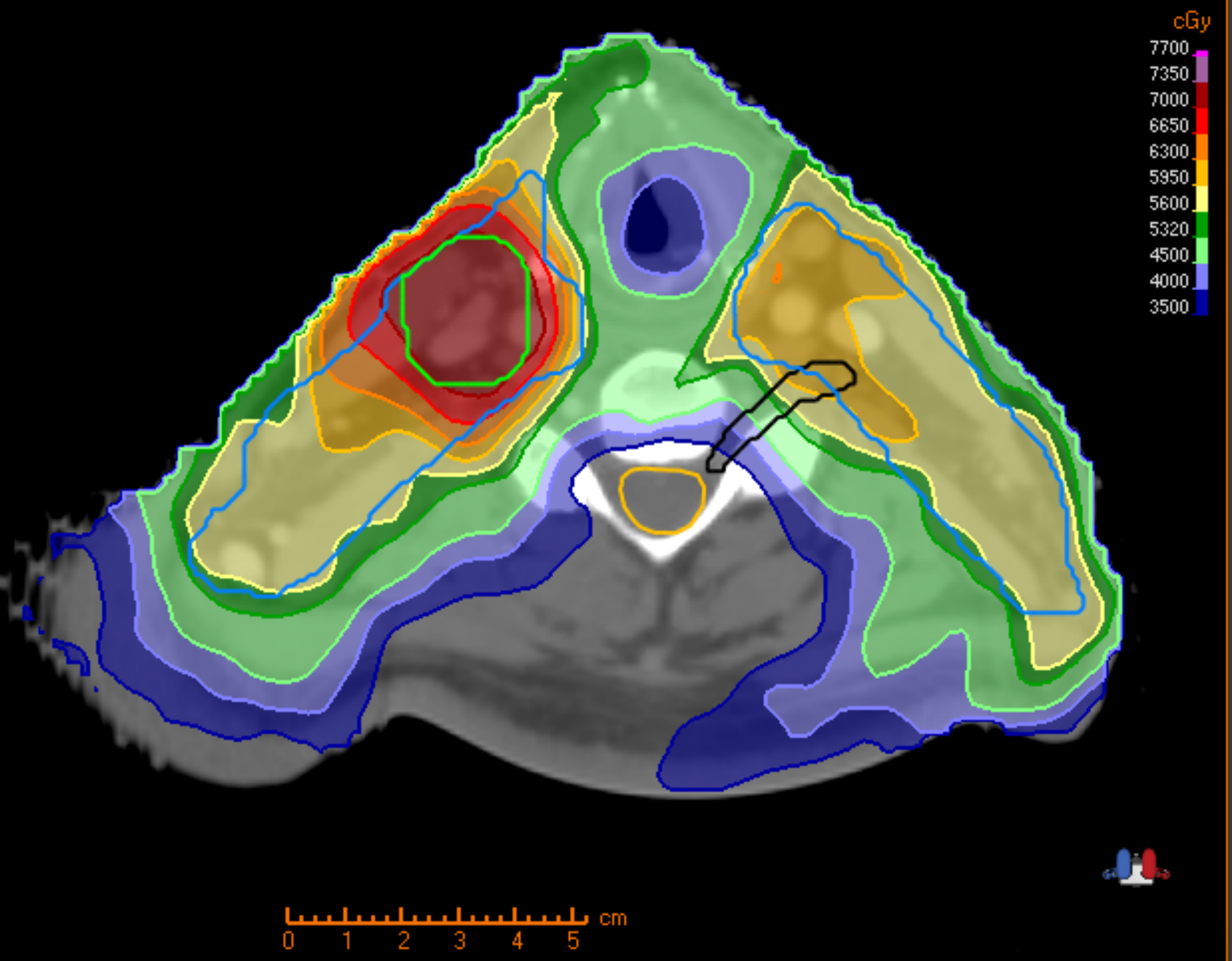}&\multirow{2}{*}[78pt]{\hspace{0pt}\includegraphics[width=0.6\linewidth,clip=]{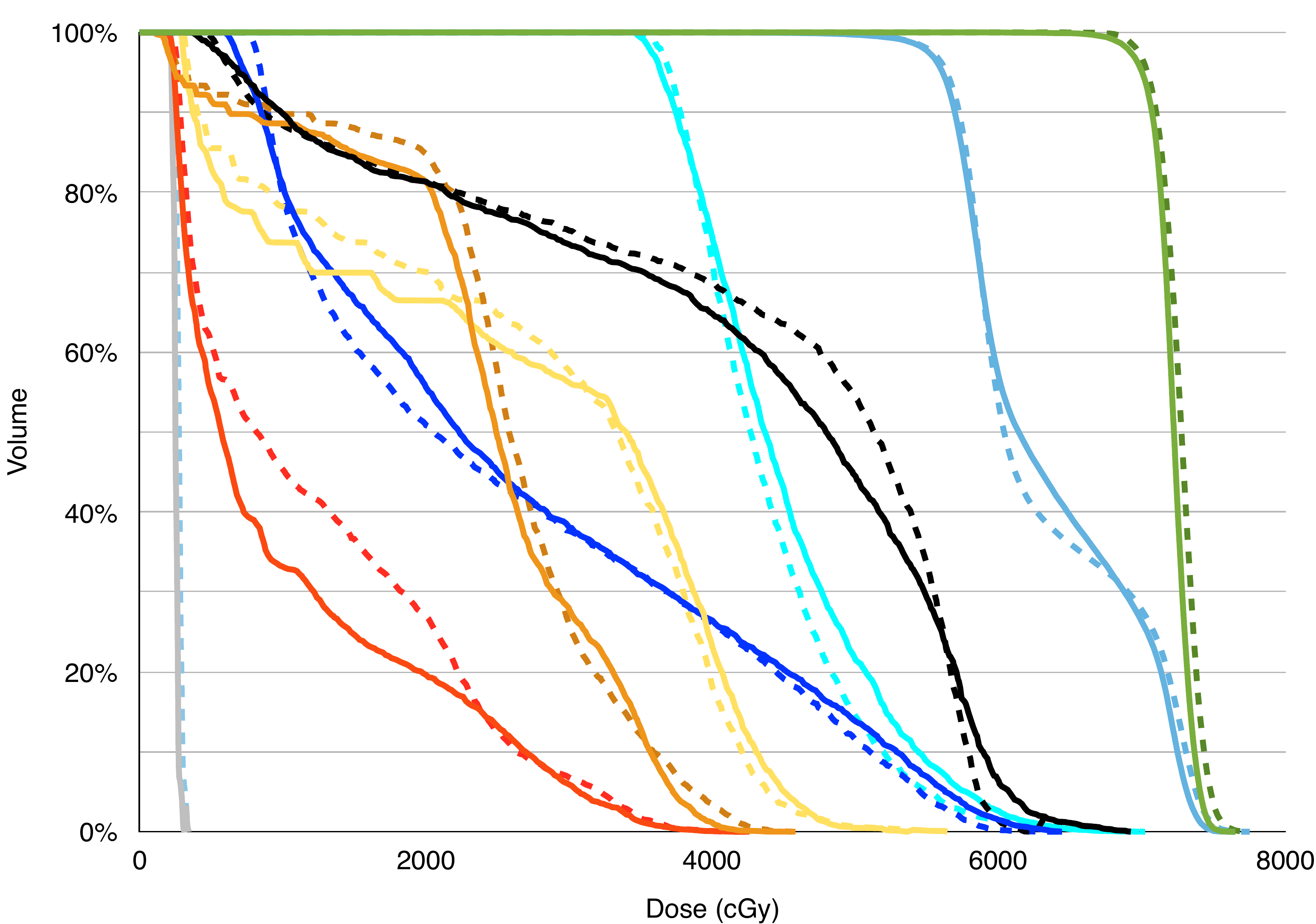}}\\
          \includegraphics[width=0.25\linewidth,clip=]{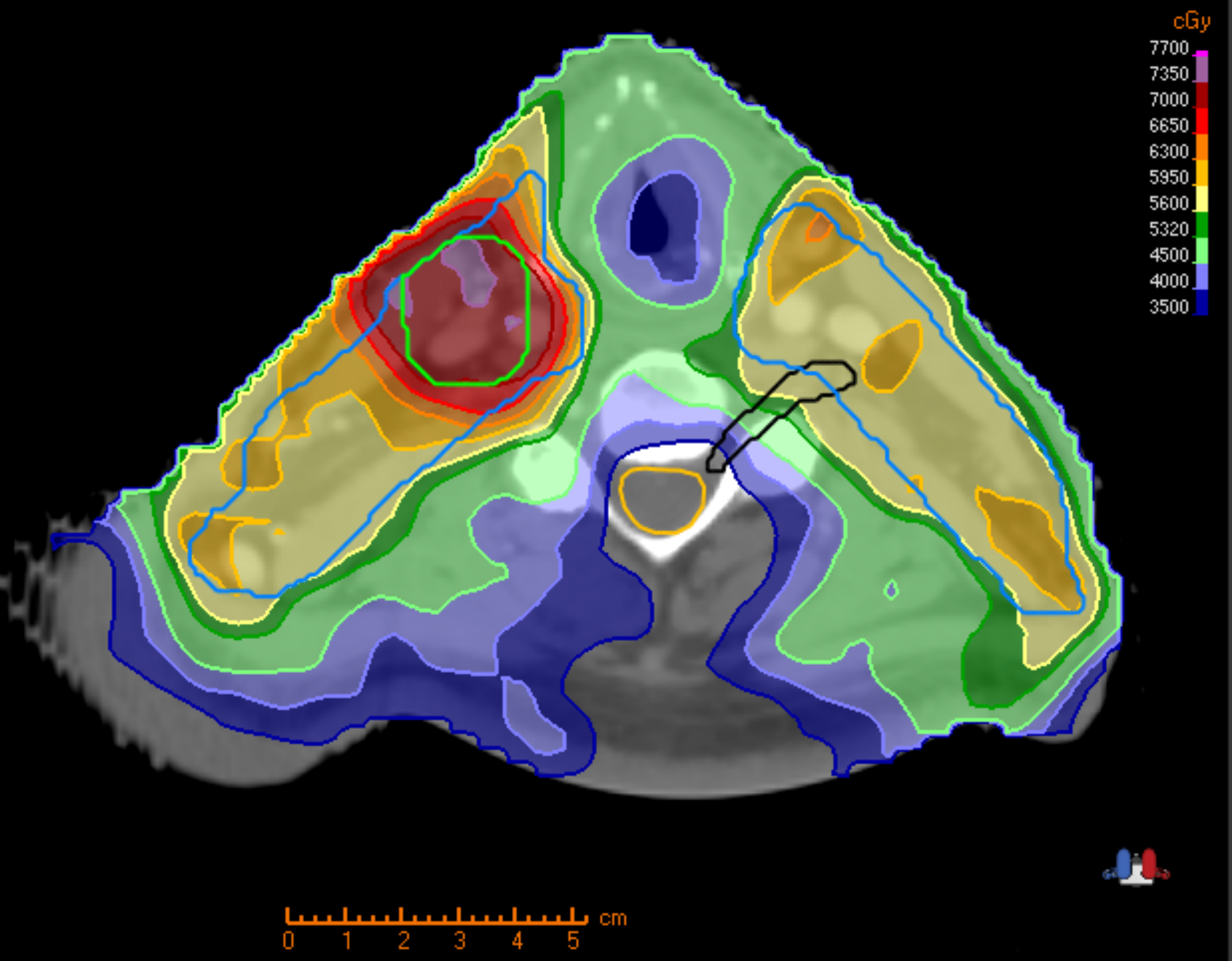}&\\
    \end{tabular}
	\end{center}
	\caption[]{Comparison of cARF-CRF[ROI] and clinical plans. Axial dose distribution for cARF-CRF[ROI] (top left) and clinical plan (bottom left) and corresponding dose volume histograms (right) for Patient 6. The dose distributions show the high risk PTV prescribed to 7000 cGy (green line), elective PTV prescribed to 5600 cGy (light blue line), and the left brachial plexus (black line). The DVHs for the cARF-CRF[ROI] (solid line) and clinical (dashed line) include the following ROIs: brainstem (red),  chiasm (grey), cord (orange), esophagus (yellow), larynx (cyan), left parotid (blue), left brachial plexus (black), elective PTV (light blue), and high risk PTV (green). The agreement between the CRF-ARF[ROI] and clinical plan DVHs is good. Patient 6 was the only patient in which the cARF-CRF[ROI] achieved less dose evaluation criteria than clinical. This is shown by the higher maximum dose for the left brachial plexus, in which the cARF-CRF[ROI] exceeds the criteria limit. }
    \label{fig::DVH} 
\end{figure*}

Dose mimicking was able to provide excellent agreement between the original clinical dose distribution and the final clinical plan dose distribution. The overall mean absolute dosimetric discrepancy between the pre- and post- mimicked dose distributions across all patients and for each metric evaluated was 111 cGy $\pm$ 184 cGy. In addition, the clinical plans showed the lowest number of overall dose evaluation criteria failing as a result of dose mimicking with only two original clinical dose evaluation criteria passing and subsequently failing due to the dose mimicking process. At the same time, dose mimicking improved the clinical plan dose distribution to achieve six more dose evaluation criteria than the original clinical dose distribution; result not included in table. Therefore, there was a net increase of four criteria passed between original and the final plan dose distributions for clinical overall with dose mimicking always resulting in the same or more dose evaluation criteria being achieved for each patient (\tabref{tb::resultSummary}).

The results presented here show cARF-CRF[ROI] plans provide excellent agreement and often show improvement over clinical plans when evaluating the dose against the dose evaluation criteria (\tabref{tb::resultSummary}). Overall, the cARF-CRF[ROI] plan achieved more criteria in seven patients, an equal number in four patients and less criteria in one patient compared with the clinical plan (Patient 6, \tabref{tb::resultSummary}). For Patient 6, the original clinical and predicted cARF-CRF[ROI] dose distributions achieved the same number of criteria, however dose mimicking was able to improve the clinical plan dose distribution and, consequently, improve two criteria that were just barely failing the limit to instead achieve the criteria. We discuss this patient in detail later in this section. 

There were three cases in which the clinical plan achieved an OAR clinical criteria that the cARF-CRF[ROI] plan did not (\figref{fig::CriteriaDiff}). These included the left parotid (Patient 11), mandible (Patient 4) and left brachial plexus (Patient 6). The left brachial plexus and mandible ROIs are not used by the cARF methods and therefore including features from  these ROIs in the CRF may improve these results. In the case of the left parotid, the CRF clearly predicted a higher dose than was actually required, which indicates that a different patient from the atlas would have been more appropriate.

Though not necessarily achievable dose distributions, Shiraishi and Moore did observe increase sparing in their voxel-based predicted dose distributions. We observed a similar trend in our mimicked cARF-CRF[ROI] plans (\figref{fig::DoseDiff}); achieving increased coverage in targets (52 versus 44) and increased sparing for OARs (41 versus 31).

Of the proposed cARF methods, only the cARF-CRF[ROI] method actually achieved more dose evaluation criteria after dose mimicking (\tabref{tb::resultSummary}). This result suggests that the cARF-CRF[ROI] method is able to predict more realistic dose distributions consistent with clinical dose distributions and therefore will have high agreement between pre- and post- mimicking (\figref{fig::PrePostMimic}). The cARF-CRF[ROI] method incorporates OAR features for atlas selection and also uses the CRF to better distribute the spatial dose within ROIs. The methodology can directly convey all the objectives and associated compromises as a single spatial dose objective. 

The other cARF methods achieve fewer criteria after mimicking and are clearly predicting dose distributions which cannot be as well realized in practice. Note that while it would be straightforward to generate predicted dose distributions that meet all dose evaluation criteria, the resulting distributions would be unrealistic and not achievable. For cases in which there are competing priorities, not all of the dose evaluation criteria can simultaneously be achieved and trade-offs between treating the target and sparing OARs must be reconciled as is common in head and neck. The strength of the cARF method is that these trade-offs can be learned and the spatial dose objective implicitly determines where in the plan the trade-offs are required and what are the achievable doses to best manage these trade-offs. As shown, the cARF-CRF[ROI] generates predicted dose distributions that are the most realistic in terms of managing trade-offs and therefore are also more accurately mimicked. As trade-offs are predicted in a less realistic fashion with the other cARF others, dose mimicking sees these voxels as outlier data-points, and in trying to mimic the dose to these voxels, the overall fit worsens to all of the other voxels. 

Comparing our method with and without ROI features indicates interesting results with a degree of over-fitting, perhaps due to contouring variation between patients. Theoretically, estimating $P(d_{\pixel}|F_{\pixel})$ and the associated atlas- selection probabilities should be easier given additional ROI features. The cARF-CRF predicted dose distribution i.e without OAR ROI features, achieves the same number of criteria in 8 patients and achieves more criteria in 4 patients in comparison to the cARF-CRF[ROI] predicted dose distribution (\tabref{tb::resultSummary}). However, following dose mimicking, there is a reversal with the cARF-CRF[ROI] achieving more criteria in 6 patients and less criteria in one patient. This implies that the ROIs features enabled some over-fitting to the atlas-selection data for the predicting the prior. Examining \figref{fig::featureImp} yields insight into why. While gradient features are less precise for spatially localizing dose, and hence harder to mimic the resulting plans, they encode information about not only the shape, but the image appearance within a region. As a result the algorithm can pick better atlases without ROI features, but is less equipped to accurately map the dose distribution from those atlases onto a novel patient in an achievable manner from a dose mimicking perspective.

It is interesting to note that the cARF method only has the target ROIs as an input, and the cARF[ROI], cARF-CRF, and cARF-CRF[ROI] methods only require the limited number of ROIs defined in \secref{sec::doseMimicking}. Despite this limitation the cARF method still achieved a reasonable number of criteria, with three plans achieving more criteria, three the same, and six fewer in comparison to clinical. 

As noted above, the only case in which the clinical plan achieved more criteria than the cARF-CRF[ROI] was Patient 6 (\figref{fig::DVH}), which was also the only case in which the clinical plan achieved all the dose evaluation criteria. The other cARF methods also performed better than the cARF-CRF[ROI] method for Patient 6, which was the only patient to show this result. The primary difference between the plans for patient 6 was that the cARF-CRF[ROI] method had a much higher dose than required to the left brachial plexus. The maximum dose was 6900 cGy, which far exceeded the 6300 cGy limit. The maximum dose to the left brachial plexus for the other plans were: 6249 cGy (clinical), 6293 (cARF-CRF), 5991 cGy (cARF[ROI]), and 6007 cGy (cARF). Interestingly, for the five atlas patients that were selected for the CRF dose prior optimization for this patient, the maximum doses to the left brachial plexus were 5953 cGy, 5924 cGy, 6862 cGy, 6458 cGy, and 6041 cGy, respectively. Therefore, two of the atlas patients exceed the dose evaluation criteria limit of 6300 cGy. The dose prior is an average over the most similar atlases and is distorted by the maximum doses of 6862 cGy and 6458 cGy. Without those two atlases the prior constraints in the CRF would have more strictly limited the dose to the targets, and thus lowered the left brachial plexus dose. The cARF-CRF achieved this criteria. It did so by selecting a different set of atlases which did not include those higher dose ranges as reasonably probable. 

There are three potential reasons for this result: the majority of the patients in the database do have some competing priorities and most patients do not achieve all of the ideal criteria; there may be over-fitting between the ROI and non-ROI versions of the cARF-CRF methods (as discussed previously);  and the left brachial plexus is not an input into the cARF or CRF and therefore there are no explicit left brachial plexus features used for training or testing. The potential solutions include adding more training data, better curating of the training database to remove sub-optimal training dose distributions, and including left brachial plexus features. For example, including more patients that achieve all criteria in the training database, although clinical practice may not dictate this. 
  
In this work we used the clinical beams orientations to focus the comparison between the clinical inverse planning dose-volume objective based pipeline and voxel-based dose prediction pipeline. This is not a technical limitation of the automated pipeline as the beam orientations from the atlas patients selected for the cARF part of the pipeline can readily be extracted from their corresponding treatment plans and used to fulfill this requirement. In head and neck this is a minor point of variation as the number and orientation of beams is highly standardized across all of the training and testing patients. For treatment sites in which the beam orientation varies considerably between patient and/or for which the beam orientation impacts the dose distribution, beam orientation selection would be a consideration and is an area of future investigation. 

Our method is a knowledge-based approached, and thus first requires a training database for its application to a new treatment site. We previously validated that, intuitively, the number of required training atlases is a function of the variation and complexity of a given site \cite{mcIntosh2016TMIAutoPlan}. Our preliminary results here are promising, with only 54 training patients proving sufficient to accurately plan a complex site with many ROI intersections that is a time consuming manual process. It is important to note though that the method is easily amenable to adding new training atlases without retraining the entire system, and therefore can be deployed with a small atlas set and expanded over time as more plans are approved. The number of training patients required for other treatment sites and further testing the applicability of the method to include more testing cases and additional treatment sites are future areas of investigation.

\section{Conclusions}
\label{sec::conclusions}

This paper presents the first evaluation of achievable dose distributions from knowledge-based per-voxel dose prediction using machine learning in a fully automated treatment planning framework. The cARF methods presented only require the images, target delineations, and a limited number of anatomical OARs as an input. The method generates a spatial dose objective, which eliminates the need for specify dose-volume objectives in the canoncial planning process. Based on overall scoring dose evaluation criteria, the best automated plans were generated using the cARF-cRF[ROI] method, in which we were able to achieve more criteria and provide superior target coverage and OAR sparing compared to clinical plans in 12 head and neck patients. In addition, our results demonstrate that while the purely voxel-based cARF and non-CRF methods produce reasonable plans on import, these dose distributions are more difficult to mimic than the dose distributions of the cARF-CRF[ROI] method. Therefore, the cARF-CRF[ROI] can predict spatially accurate dose distributions that are readily mimicked to generate complete treatment plans. The use of feature-based machine learning and atlas-selection makes the method highly generalizable to other treatment sites and treatment modalities.

\bibliographystyle{dcu}
\bibliography{autoPlan}

\end{document}